\documentclass[twoside,onecolumn]{article}
\usepackage{slashed}
\usepackage{graphicx}
\usepackage{subfigure}
\usepackage[usenames, dvipsnames]{color}
\usepackage{graphics}
\usepackage{hyperref}
\usepackage{bm}
\usepackage{amsmath}
\usepackage{color}
\usepackage{amsfonts}
\hypersetup{backref,
colorlinks=true,
linkcolor=cyan,
linktoc=page,
citecolor=cyan, 
urlcolor=cyan}
\usepackage{xcolor}
\usepackage[margin=1in]{geometry}

\everymath{\displaystyle} 
\newcommand{\pr}[1]{\ensuremath{\left[#1\right]}} 
\newcommand{\pc}[1]{\ensuremath{\left(#1\right)}} 
\newcommand{\md}[1]{\ensuremath{\left\vert#1\right\vert}} 
\newcommand{\av}[1]{\ensuremath{\left\langle#1\right\rangle}} 
\let\oldAA\AA
\renewcommand{\AA}{\text{\normalfont\oldAA}}

\DeclareMathOperator{\hH}{\hat{H}}

\DeclareMathOperator{\hb}{\hat{b}}
\DeclareMathOperator{\he}{\hat{e}}

\DeclareMathOperator{\hp}{\hat{p}}
\DeclareMathOperator{\hP}{\hat{P}}
\DeclareMathOperator{\hr}{\hat{r}}
\DeclareMathOperator{\hq}{\hat{q}}
\DeclareMathOperator{\hh}{\hat{h}}

\title{Quantum Theory of Multisubband Plasmon-Phonon Coupling}

\author{%
Sofia Ribeiro \thanks{Current address: Joint Quantum Centre (JQC) Durham-Newcastle, Department of Physics, Rochester Building, Durham University, United Kingdom} \\[1ex] 
\normalsize Laboratoire Mat\'eriaux et Ph\'enom\`enes Quantiques,\\
\normalsize UMR7162, Universit\'e de Paris, F-75013 Paris, France \\
\and 
Angela Vasanelli, Yanko Todorov and Carlo Sirtori,\\[1ex]
\normalsize Laboratoire de Physique de l'\'Ecole Normale Sup\'erieure, ENS,\\
\normalsize Universit\'e PSL, CNRS, Sorbonne Universit\'e, \\
\normalsize Universit\'e de Paris, F-75005 Paris, France
}
\begin{document}
\maketitle

\begin{abstract}
We present a theoretical description of the coupling between longitudinal optical phonons and collective excitations of a two-dimensional electron gas. By diagonalizing the Hamiltonian of the system, including Coulomb electron--electron and Fr\"ohlich interactions, we observe the formation of multisubband polarons, mixed states partially phonon and partially multisubband plasmon, characterized by a coupling energy which is a significant fraction, up to $\sim$ $40 \%$, of the phonon energy. We demonstrate that multisubband plasmons and longitudinal optical phonons are in the ultra-strong coupling regime in several III--V and II--VI material systems.\end{abstract}

\section{Introduction}

Longitudinal optical (LO) phonons provide an important scattering channel for a two-dimensional electron gas confined in a semiconductor quantum well. The~effect of this coupling determines the operation of mid-infrared quantum optoelectronic devices, such as quantum cascade lasers~\cite{QCL} and detectors~\cite{QCD}, as~it limits the lifetime of carriers in excited subbands~\cite{PRB40_1074(1989)}. In~order to take into account the scattering between the electron gas and the phonons in the device operation, a~single-particle picture is usually employed: even though a multitude of electrons is involved in the device operation, their density is usually sufficiently low to neglect collective~phenomena. 

A cooperative behaviour of the two-dimensional electron gas, due to the electron--electron interactions (Coulomb interactions), has been observed in the infrared spectrum of highly doped semiconductor quantum wells~\cite{PRL109_246808(2012), RMP54_437(1982), PRB85_045304(2012)}. If~a single electronic subband is occupied, one signature of this collective behaviour is the so-called plasma shift: The absorption spectrum presents a resonance which is shifted towards higher energies as compared to that of the transition between the ground and first excited subband. This resonance corresponds to the excitation of a collective mode of the system, known as {intersubband plasmon} \cite{PSSb177_9(1993), RMP54_437(1982)}. In~the case where two subbands are occupied, the~dipole-dipole coupling between intersubband (ISB) plasmons associated with the optically active electronic transitions leads to a strong renormalization of the absorption spectrum~\cite{warburton, APL102_031102(2013)}: The energies of the peaks are shifted with respect to the single particle transitions and the oscillator strength is redistributed in the high energy mode. If~several subbands are occupied, the~entire oscillator strength is concentrated into a single resonance, corresponding to a many-body excitation of the system, the~{multisubband plasmon}, in~which the dipole-dipole Coulomb interaction locks in phase all the allowed transitions between confined states~\cite{PRL109_246808(2012), PRB90_035305(2014),PRB86_125314(2012), PRB90_115311(2014)}. 
The multisubband (MSB) plasmon is a charge density wave where the collective dipole oscillates along the growth direction of the quantum well, while the plasmon propagates in the quantum well plane, with~a characteristic in-plane wavevector. 
The MSB plasmon has a superradiant nature~\cite{PRL115_187402(2015)}: This means that its spontaneous emission lifetime depends on the number of electrons involved in the interaction with light and it can be even shorter than the typical non-radiative lifetime~\cite{APL107_241112(2015)}. This results in a radiative broadening of the MSB plasmon~spectra. 

\begin{figure}[t!]
\centering
\includegraphics[width=0.55\linewidth]{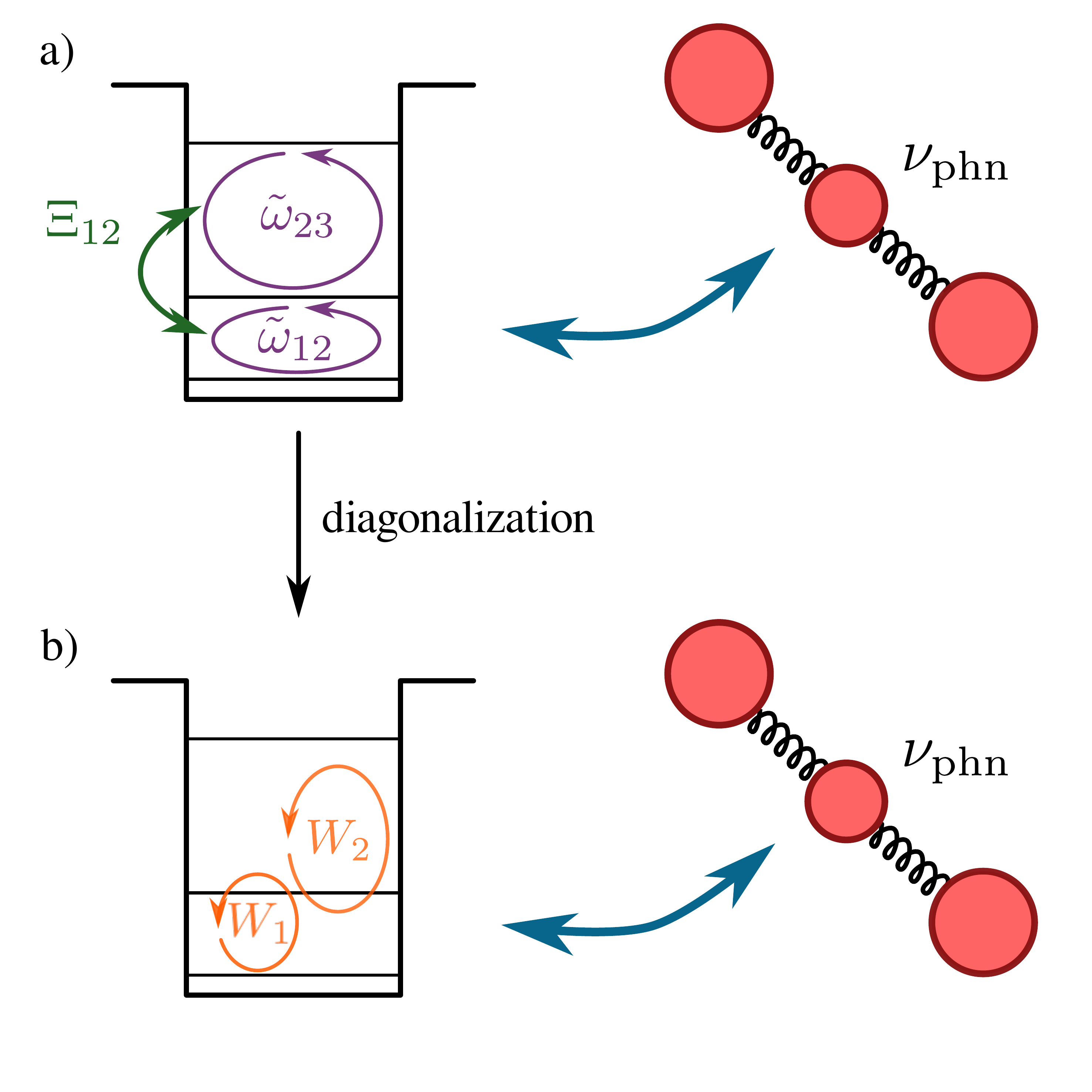}\\
\caption{(Color online) (\textbf{a}) Scheme of the intersubband plasmons (purple arrows) and the interactions between them (green arrow) and their interaction with phonons (blue arrow). 
(\textbf{b}) Scheme of the multisubband plasmons of the system after diagonalization of intersubband interactions (orange arrows) and their interaction with phonons (blue arrow). \label{Fig:MSP-Phonon}}
\end{figure}

In the last years there has been a strong research activity on the realization of quantum optoelectronic devices whose operation is determined by the properties of intersubband or multisubband plasmons~\cite{PRX5_011031(2015), manceau_APL, PRB97_075402(2018),delteil_PRB}. The~interaction of these quasi-particles with optical phonons is of paramount importance for the operation of the device. 
De Liberato and Ciuti have first studied the possibility of achieving stimulated scattering and lasing of intersubband cavity polaritons, the~quasi-particles issued from the strong coupling between intersubband plasmons and a cavity mode~\cite{PRL102_136403(2009)}. Their proposal exploits the relaxation from the upper to the lower polariton branch by means of scattering by longitudinal optical (LO) phonons. Delteil~et~al.~\cite{delteil_PRB} have provided experimental evidence of scattering between intersubband polaritons and LO-phonons in an electroluminescent device. Following these works, the~possibility of an intersubband polariton optically pumped laser has been investigated~\cite{PRX5_011031(2015), manceau_APL}. The~proposed device is based on scattering between LO-phonons and intersubband polaritons with a dispersion designed to favour polariton condensation. All these works concern mid-infrared devices based on intersubband polaritons interacting with longitudinal optical phonons. The~strength of the polariton-phonon interaction is in this case proportional to the Hopfield coefficient related to the electronic part of the initial and final polariton state. It can thus be interesting to study this scattering process in the absence of a cavity mode, in~a device based on multisubband plasmons coupled with free space radiation~\cite{PRL115_187402(2015), APL107_241112(2015)}.
While the interaction between these collective electronic modes and the electromagnetic field has been widely investigated, a~complete description of their coupling with the different phonon modes is still missing. 
In reference~\cite{PRB85_125302(2012)}, a~microscopic quantum theory of ISB polarons, issued from the coupling between an ISB plasmon and a longitudinal optical phonon, is presented. The~authors showed, for~different materials, that the coupling between the phonons and the ISB transitions can be very strong, thanks both to the collective nature of the ISB excitations and to the natural tight confinement of optical phonons. In~this paper, we extend the microscopic theory developed in the previous study to consider MSB polarons, that is, a~theory of MSB excitations coupled to LO phonons in semiconductor quantum wells (see Figure~\ref{Fig:MSP-Phonon}). For~such, we will consider an infinite quantum well with more than one populated subband. Using a second quantization formalism, we will reduce the full electron--phonon Hamiltonian to a quadratic, bosonic form, from~which we then calculate the MSB polaron dispersions. We will show that the coupling energy between MSB excitations and phonons is a significant fraction of the phonon energy, resulting in the achievement of the ultra-strong coupling regime between MSB excitations and longitudinal optical phonons.

This paper is organized as follow: We begin by developing the general theory of the coupling between MSB transitions and LO phonons, in~Section~\ref{Sec:theory}. In~this section, we consider the Hamiltonian for free electrons and Coulomb interactions which we broaden to find a bosonic Hamiltonian accounting for ISB transitions. We extend from ISB transitions to ISB plasmons and finally, by~considering the interactions between them, we diagonalize the Hamiltonian to find MSB plasmons. After~finding the interaction Hamiltonian between MSB plasmon and LO phonons, we diagonalize the full interaction Hamiltonian in Section~\ref{Sec:diag}. In~Section~\ref{Sec:results}, we apply the results to the case of an infinite quantum well, showing when and how the strong-coupling regime can be reached. Finally, in~Section~\ref{Sec:conclusions}, we give concluding~remarks.
%
\section{Theoretical~Background \label{Sec:theory}}

In this section, we will develop a microscopic theory of MSB polarons in a quantum well for zero temperature.
We study the coupling of MSB excitations with longitudinal optical phonons, considering also the role of Coulomb electron-electron interaction in order to derive the MSB plasmon-phonon coupling. The~total Hamiltonian of the system is given by:
\begin{align}
\hH= \hH_\text{e} + \hH_\text{phn} + \hH_\text{e-e} + \hH_\text{e-phn},
\end{align}
with $\hH_\text{e}$ the bare electron Hamiltonian, $\hH_\text{phn}$ the bare phonon Hamiltonian, $\hH_\text{e-e}$ describes the electron-electron (Coulomb) interaction and $\hH_\text{e-phn}$ the electron-phonon~coupling.

\subsection{Free Electron and Electron-Electron Coupling~Hamiltonian}

The single-particle Hamiltonian omitting the electron spin index (all interactions we considered are spin conserving) is given by
\begin{align}
\hH_\text{e} = \sum_{i \mathbf{k}} \hbar \omega_{i \mathbf{k}} \he_{i \mathbf{k}}^\dagger \he_{i \mathbf{k}}
\end{align}
with $\he_{i \mathbf{k}}^\dagger, \, \he_{i \mathbf{k}}$ are the fermionic creation and destruction operators for an electron in the subband $i$.
Since we are interested in studying the ISB transitions, transitions between the different electronic states, we define the operator $\hat{B}^\dagger_{ji\mathbf{q}} = \sum_\mathbf{k} \he^\dagger_{j \mathbf{k+q}} \he_{i \mathbf{k}}$ \cite{PRB85_045304(2012)}. We find
\begin{align*}
\pr{\hat{B}^\dagger_{ji\mathbf{q}}, \hH_e} = - \sum_\mathbf{k} \hbar \pc{\omega_{j \mathbf{k+q}} - \omega_{i \mathbf{k}}}  \he^\dagger_{j \mathbf{k+q}} \he_{i \mathbf{k}}.
\end{align*}

In the long-wavelength limit, the~excitation wave vector $\mathbf{q}$ is small compared to the typical electron wave vectors $\mathbf{k}$ and one can consider $\omega_{j \mathbf{k+q}} - \omega_{i \mathbf{k}} \approx \omega_j - \omega_i \equiv \omega_{ji}$,
then
\begin{align*}
\pr{\hat{B}^\dagger_{ji\mathbf{q}}, \hH_e} = - \hbar \omega_{ji} \hat{B}^\dagger_{ji\mathbf{q}}.
\end{align*}

Our goal is then to replace the fermionic Hamiltonian $\hH_\text{e}$ by an effective bosonic Hamiltonian describing the transitions between subbands. To~carry out this approach, we need to replace the operators $\hat{B}$ by effective bosonic operators. For~this purpose, we compute the commutation relations
\begin{align*}
\pr{\hat{B}_{ji\mathbf{q}},\hat{B}^\dagger_{ji\mathbf{q}} } = \sum_\mathbf{k} \pc{\he^\dagger_{j \mathbf{k}} \he_{j \mathbf{k}}-\he^\dagger_{i \mathbf{k+q}} \he_{i \mathbf{k+q}} } = \hat{N}_j - \hat{N}_i.
\end{align*}

Having this result in mind it is possible to define the normalized operators through the relation
\begin{align*}
\hat{B}^\dagger_{ji\mathbf{q}} = \sqrt{\Delta N_{ji}} \hb^\dagger_{ji \mathbf{q}}
\end{align*}
with $\Delta N_\alpha = \av{\hat{N}_j} -\av{\hat{N}_i}$ the difference in the subband populations. 
Labeling the electronic transitions $\alpha \equiv i\to j$, we introduce the operators describing ISB transitions as
\begin{subequations}
\begin{eqnarray}
\hb^\dagger_{\alpha \mathbf{q}} &= \frac{1}{\sqrt{\Delta N_\alpha}} \sum_\mathbf{k} \he^\dagger_{j\mathbf{k+q}} \he_{i \mathbf{k}}, \\
\hb_{\alpha \mathbf{q}} &= \frac{1}{\sqrt{\Delta N_\alpha}} \sum_\mathbf{k} \he^\dagger_{i\mathbf{k}} \he_{j \mathbf{k+q}}.
\end{eqnarray}
\label{eq:boperators}
\end{subequations}

In this formalism, we can finally redefine $H_\text{e}$ as an effective bosonic Hamiltonian
\begin{align}
\hH_\text{e} = \sum_{\alpha \mathbf{q}} \hbar \omega_\alpha \hb^\dagger_{\alpha \mathbf{q}} \hb_{\alpha \mathbf{q}}.
\end{align}
$\hb^\dagger_{\alpha \mathbf{q}}, \, \hb_{\alpha \mathbf{q}}$ are bosonic operators in the limit of weakly excited systems $\pr{\hb_{\alpha \mathbf{q}}, \hb^\dagger_{\alpha \mathbf{q}}}= \delta_{\mathbf{q,q'}}$ \cite{PRL102_136403(2009)}.

Up to now the electron--electron or Coulomb interactions have not been considered. However, for~high electron densities, many-body effects cannot be neglected. The~main effect of the electron--electron interaction is to give a collective character to the elementary excitations, namely creating the so-called {intersubband plasmons}. 
Let us start then by considering two electrons in the subband $j$ and $n$ with momenta $\mathbf{k}$ and $\mathbf{k'}$ that are scattered into subbands $i$ and $m$ with momenta $\mathbf{k+q}$ and $\mathbf{k'+q}$, respectively.
In order to treat the Coulomb electron--electron interaction, we start by the second quantized form of the Hamiltonian describing the Coulomb interaction~\cite{PRL79_4633(1997)}
\begin{align}
\hH_\text{e-e} = \frac{1}{2} \sum_{i,j,m,n} \sum_{\mathbf{q,k,k'}}  V_{imnj,\mathbf{q}} \he^\dagger_{i, \mathbf{k+q}} \he^\dagger_{m, \mathbf{k'-q}} \he_{n, \mathbf{k'}} \he_{j, \mathbf{k}}.
\end{align}

Here, the~factor $1/2$ accounts for the double count of the particles. The~two-dimensional Coulomb matrix element is given by
\begin{align}
V_{imnj, \mathbf{q}} &= \frac{V_\mathbf{q}}{S} \int dz \int dz' \psi_i (z) \psi_j (z)e^{-q \md{z-z'}} \psi_m (z') \psi_n (z') 
\label{eq:formfactor}
\end{align}
where $V_\mathbf{q}$ is the Fourier transform of the Coulomb potential in two dimensions defined by 
\begin{align}
V_\mathbf{q} = \frac{e^2}{2 \varepsilon_0 \varepsilon_\infty \md{\mathbf{q}}}.
\end{align}

Note that, following ref.~\cite{PRB85_125302(2012)}, we used the high-frequency dielectric constant $\varepsilon_\infty$ instead of the static one. This is due to the fact that $\varepsilon_s$ includes the effect of the coupling to LO phonons, which are already treated exactly in the~Hamiltonian. 

The strength of the electron--phonon interaction, as~we will see in the next section, will be determined by the same integral as the Coulomb matrix element:
\begin{align}
I_{imnj} (q)  = \int \, dz \, \int \, dz'  \psi_{i} (z) \psi_{j} (z) e^{-q \md{z-z'}} \psi_{m}(z') \psi_{n} (z')
\end{align}

For electron--phonon interaction, only cases where $j \to i = n \to m$, with~$j \neq i$, are considered. However, for~Coulomb interaction other transitions can be considered.
In a symmetric well, the~symmetry of the wave functions results in $I_{imnj} (q) \neq0$ only when the sum of the individual indices $j, \,n, \, i$ and $m$ is even~\cite{PRB59_15796(1999)}. For~an infinite quantum well these integrals will be evaluated analytically later. 
Matrix elements with different $j \to i,\; n \to m$ indices represent ISB excitations where each electron is scattered from one subband to another; these processes are responsible for the depolarization shift. The~terms other than those responsible for the depolarization shift are strongly suppressed due to their lack of collective enhancement and can be treated perturbatively~\cite{PRL79_4633(1997)}.
In~reference~\cite{PRB59_15796(1999), PRB62_15327(2000)}, it is shown that the intrasubband terms do not contribute to the screening of the ISB ones at the level of the random phase~approximation.

So we can now write the electron and electron--electron Hamiltonian in terms of ISB transitions as
\begin{align}
\hH_\text{e} &+\hH_\text{e-e} = \sum_{\alpha, \mathbf{q}} \hbar \omega_{\alpha} \hb^\dagger_{\alpha, \mathbf{q}} \hb_{\alpha, \mathbf{q}}  + \sum_{\alpha, \beta,  \mathbf{q}} \frac{e^2 \sqrt{\Delta N_{\alpha}\Delta N_{\beta}}}{4 \varepsilon_0 \varepsilon_\infty S } \frac{I_{imnj} (q)}{\md{\mathbf{q}}} \pc{\hb^\dagger_{\alpha, \mathbf{q}} + \hb_{\alpha,- \mathbf{q}}} \pc{\hb^\dagger_{\beta,-\mathbf{q}} + \hb_{\beta, \mathbf{q}}} .
\label{eq:He+Hee}
\end{align}

\subsubsection{From Intersubband Transitions to Intersubband~Plasmons}

The quadratic Hamiltonian \eqref{eq:He+Hee} can be diagonalized with the Bogoliubov transformation by introducing new bosonic operators $\hp_{\alpha,\mathbf{q}}$, which satisfy
\begin{align}
\pr{\hp_{\alpha,\mathbf{q}}, \hH_\text{e} +\hH_\text{e-e}} = \hbar \tilde{\omega}_{\alpha,\mathbf{q}} \hp_{\alpha, \mathbf{q}}.
\end{align}

Following the same procedure as in reference~\cite{PRB85_045304(2012)}, it is possible to re-write the Hamiltonian in terms of operators that create and annihilate ISB plasmons which describe the ensemble of interacting intersubband dipolar oscillators. The~diagonalization of the quadratic Hamiltonian is performed considering first $\alpha = \beta$ and later extended to include the terms where  $\alpha \neq \beta$ \cite{PRL105_196402(2010)}. The~bosonic ISB plasmon operator becomes
\begin{align}
\hp_{\alpha, \mathbf{q}} = \frac{\tilde{\omega}_{\alpha, \mathbf{q}} + \omega_{\alpha}}{2 \sqrt{\tilde{\omega}_{\alpha, \mathbf{q}} \omega_{\alpha}}} \hb_{\alpha, \mathbf{q}} +  \frac{\tilde{\omega}_{\alpha, \mathbf{q}} - \omega_{\alpha}}{2 \sqrt{\tilde{\omega}_{\alpha, \mathbf{q}} \omega_{\alpha}}} \hb^\dagger_{\alpha, \mathbf{q}} 
\end{align}
with $\tilde{\omega}_{\alpha, \mathbf{q}} = \sqrt{ \omega^2_{\alpha} +  \Theta_{\alpha, \mathbf{q}}^2}$ and
\begin{align}
\Theta_{\alpha, \mathbf{q}}^2 = \frac{e^2 \omega_{\alpha}}{ \hbar \varepsilon_0 \varepsilon_\infty} \frac{\Delta N_{\alpha}}{S} \frac{I_{\alpha \alpha}(q)}{\md{\mathbf{q}}} .
\end{align}

Note that $\Theta_{\alpha,\mathbf{-q}}=\Theta_{\alpha,\mathbf{q}}$. For~$\mathbf{q} \to 0$, $\Theta_{\alpha,\mathbf{q}} ^2$ coincides with the squared plasma frequency and the new eigenvalues have exactly the frequency of the collective mode of the two-dimensional electron gas known as the {intersubband plasmon}.

The MSB plasmon Hamiltonian can now be obtained by adding the the terms $\alpha \neq \beta$, thus including the coupling between different intersubband plasmons:
\begin{align}
\hH_\text{e} &+\hH_\text{e-e} = \sum_{\alpha, \mathbf{q}} \hbar \tilde{\omega}_{\alpha, \mathbf{q}} \hp^\dagger_{\alpha,\mathbf{q}} \hp_{\alpha,\mathbf{q}} + \hbar \sum_{\alpha \neq \beta,\mathbf{q}} \Xi_{\alpha, \beta, \mathbf{q}} \pc{\hp^\dagger_{\alpha, \mathbf{q}} + \hp_{\alpha, -\mathbf{q}}}
\pc{\hp^\dagger_{\beta,-\mathbf{q}} + \hp_{\beta,\mathbf{q}}} 
\end{align}
with
\begin{align}
\Xi_{\alpha, \beta,\mathbf{q}} = \frac{\Theta_{\alpha, \mathbf{q}} \Theta_{\beta, \mathbf{q}}}{4 \sqrt{\tilde{\omega}_{\alpha, \mathbf{q}} \tilde{\omega}_{\beta, \mathbf{q}}}}\frac{\mathcal{F}_{\alpha \beta} (q)}{\sqrt{\mathcal{F}_{\alpha \alpha}(q) \mathcal{F}_{\beta \beta} (q)}},
\label{eq:xi}
\end{align}
and
\begin{align}
\mathcal{F}_{\alpha \beta} (q) = \frac{I_{\alpha \beta} (q)}{\md{\mathbf{q}}}.
\end{align}

\subsubsection{From Intersubband Plasmons to Multisubband~Plasmons}

The MSB plasmon Hamiltonian can now be diagonalized by introducing the bosonic operators $\hP_{n,\mathbf{q}}$ that are linear combinations of the operators $\hp_{\alpha,\mathbf{q}}$ describing the ISB plasmons:
\begin{align}
\hP_{n,\mathbf{q}} = \sum_\alpha \pc{x_{n\alpha} \hp_{\alpha,\mathbf{q}} + y_{n\alpha} \hp^\dagger_{\alpha, -\mathbf{q}}},
\end{align}
with $\hH_\text{plasmon} = \sum_{n,\mathbf{q}} \hbar W_{n,\mathbf{q}} \hP_{n,\mathbf{q}}^\dagger \hP_{n,\mathbf{q}}$, where it holds
\begin{align}
\pr{\hP_{n,\mathbf{q}}, \hH_\text{plasmon}} = \hbar W_{n,\mathbf{q}} \hP_{n,\mathbf{q}}.
\end{align}

Considering the commutator
\begin{align}
\pr{\hp_{\alpha,\mathbf{q}}, \hH_\text{plasmon}} &= \hbar \tilde{\omega}_{\alpha,\mathbf{q}} \hp_{\alpha,\mathbf{q}} +  \sum_{\alpha\neq \beta,\mathbf{q}} \hbar \Xi_{\alpha \beta,\mathbf{q}} \pc{\hp^\dagger_{\beta,\mathbf{q}} + \hp_{\beta,\mathbf{q}}}
\end{align}
will lead to the eigenvalue problem $\mathbf{M} \mathbf{V}_{n,\mathbf{q}} = W_{n,\mathbf{q}} \mathbf{V}_{n,\mathbf{q}}$, with~$\mathbf{M}$ is the Hopfield matrix and $\mathbf{V}_{n,\mathbf{q}} = \pc{x_{n1}, y_{n1}, \cdots, x_{nN}, y_{nN}}^T$, where 
$\sum_i \pc{ \md{x_{ni}}^2 - \md{y_{ni}}^2 } =1$ ensures the bosonicity of the operators~\cite{PRB85_045304(2012),PRB86_125314(2012)}.

In order to have numerical results, we fix a few parameters concerning the material like the effective mass and the phonon energy $\hbar \nu_\text{phn}$. However, for~the sake of simplicity, we consider an infinite potential well of length $L_\text{QW}$. This allows us to deal with analytical calculations and to have direct indications on the role of the effective mass and phonon energy on the plasmon--phonon coupling frequency. However, our model can be extended to more realistic potentials, eventually including non-parabolicity~\cite{frucci}.

The electronic wavefunctions are given by~\cite{BookGriffiths}
$$ \psi_n (z) = \sqrt{\frac{2}{ L_\text{QW}}} \sin \frac{n \pi z }{ L_\text{QW}},$$ inside the well and zero outside. Performing the integral Equation~(8) we find~\cite{PRB59_15796(1999), PRB85_125302(2012)}
\begin{align*}
\lim_{q \to 0} I_{\alpha \alpha} (q) &\to \frac{10}{9 \pi^2} q L_\text{QW},\\
\lim_{q \to 0} I_{\beta \beta} (q) &\to \frac{26}{25 \pi ^2} q L_\text{QW},\\
\lim_{q \to 0} I_{\alpha \beta} (q) &\to \frac{1}{\pi ^2} q L_\text{QW}.
\label{eq:limintq}
\end{align*}
The electron energy at the bottom of subband $n$ is given by
\begin{align*}
E_n =  \frac{n^2 \pi^2 \hbar^2}{2 m^* L_\text{QW}^2},
\end{align*}
with population at $T=0$~K
\begin{align*}
N_n = \frac{m^* L_\text{QW}^2}{\pi \hbar}  \pc{E_F - E_n},
\end{align*}
where $E_F$ is the Fermi energy. We will set the Fermi energy such that only the first and second subbands are occupied. All the following calculations can be extended to the case of $T>0$~K, by~considering the electronic distribution in the subbands. In~this case all the possible optically active transitions contribute to the formation of the multisubband plasmon~modes.

\begin{figure}
\centering
\includegraphics[width=0.55\linewidth]{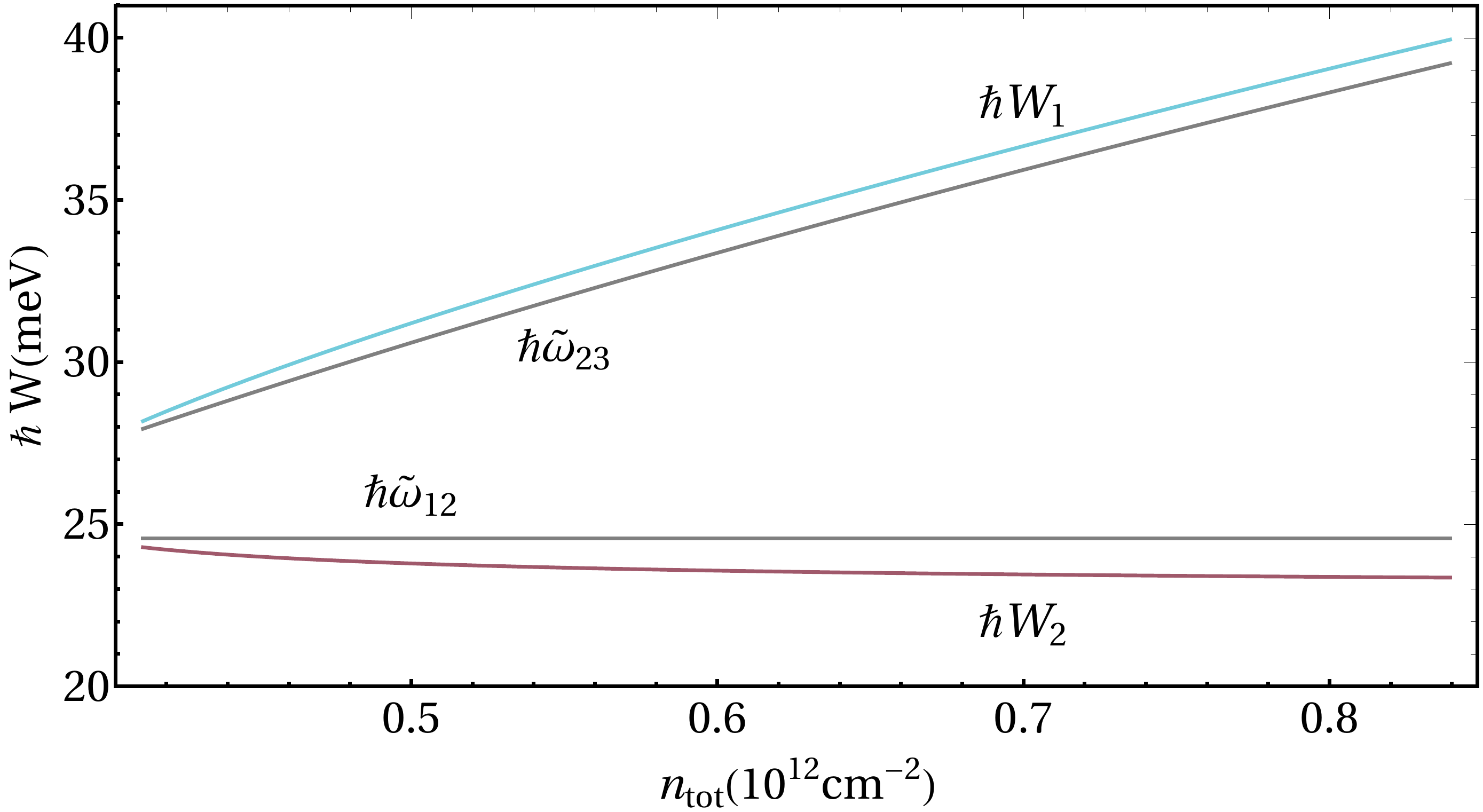}\\
\includegraphics[width=0.55\linewidth]{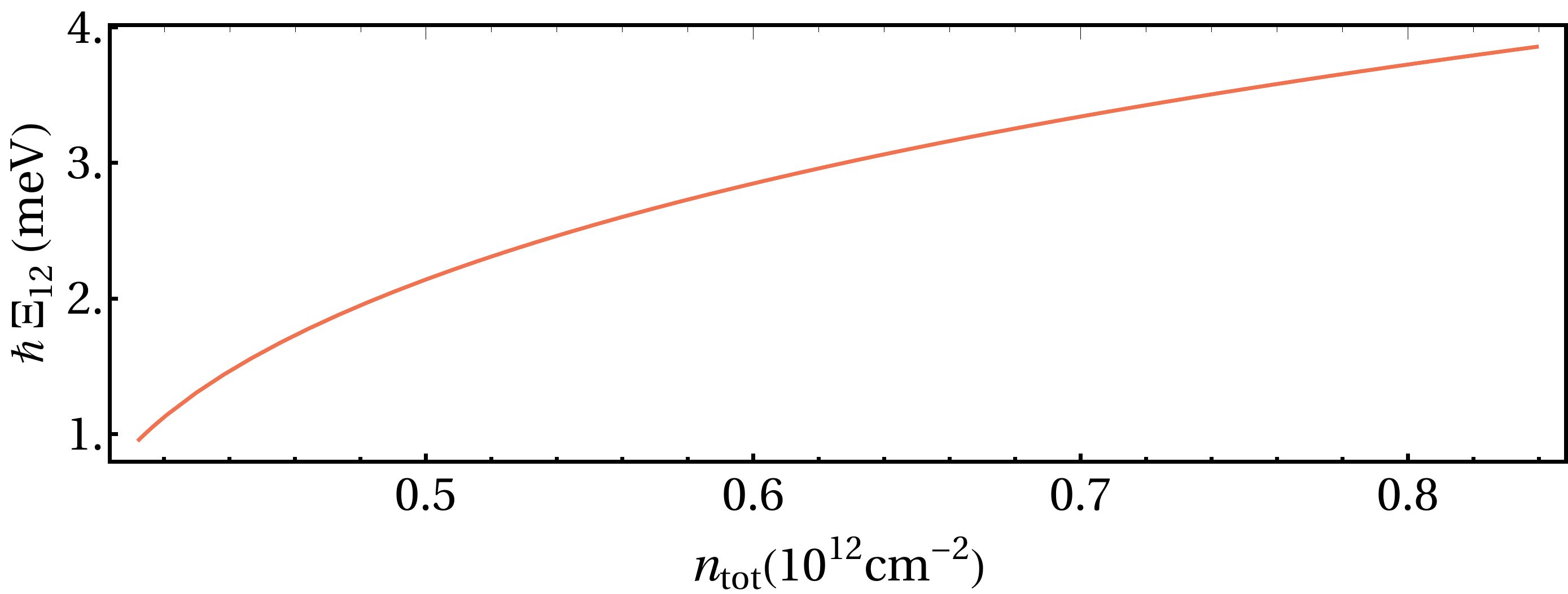}\\
\caption{(Color online) \textbf{Top}: Bare intersubband plasmon energy without interactions ($\hbar \tilde{\omega}_{12}$ and $\hbar \tilde{\omega}_{23}$) and coupled eigenmodes ($\hbar W_{1,2}$, in~purple and blue) as a function of the total doping density $n_\text{tot}$. \textbf{Bottom}: Coupling strength between the two individual intersubband plasmons as a function of the total electronic~density. \label{Fig:MSBexample}}
\end{figure}

As an example, in~Figure~\ref{Fig:MSBexample}, we show the numerical results for a quantum well of thickness $L_{QW}=$ 400~$\AA$ considering the limit $\mathbf{q}\to 0$. For~the electron effective mass in the quantum well we set $m^*=0.043$, corresponding to a GaInAs quantum well. The~bare electronic transitions have energies $\hbar \omega_{12} = 16 $~meV and $\hbar \omega_{23} = 27$~meV. The~total electronic density per unit surface $\Delta n_\text{tot}$ is variable and, as~such, also $\hbar \tilde{\omega}_{12}$. The~role of the collective effects is evident when we compare the energies of the coupled states $\hbar W_{1,2}$ with those of the uncoupled ISB plasmons, $\hbar \tilde{\omega}_{12}$ and $\hbar \tilde{\omega}_{23}$.
\unskip

\subsection{Interaction between Phonons and Intersubband~Excitations}

{{In this section we describe the model used to treat the electron--phonon interaction. As~we are interested in the case of a plasmon frequency close to the LO phonon one, we neglect confinement effects on the phonons and consider bulk values for their frequencies~\cite{PRB85_125302(2012), arora}. Furthermore, we consider quantum wells of much larger thickness than the lattice constant. As~a consequence, we neglect interface phonon modes, whose amplitude decreases exponentially away from the interface~\cite{mori, fomin}. The~phonon dispersion is also neglected because we consider only phonons with small in-plane wave vectors. The~free phonon Hamiltonian is thus described by means of boson operators $\hh^\dagger_{\mathbf{q}, q_z}, \hh_{\mathbf{q}, q_z}$, where the operator can be thought of as creating a phonon excitation traveling with in-plane $\mathbf{q}$ and out-of-plane wave vectors $q_z$,}}
\begin{align}
\hH_\text{phn} = \sum_{\mathbf{q}, q_z} \hbar \nu_\text{phn} \hh^\dagger_{\mathbf{q}, q_z} \hh_{\mathbf{q}, q_z}. 
\end{align}

{{We consider the coupling between intersubband electronic excitations and phonon modes.}}
Each electronic transition $\alpha$ will couple to multiple phonon modes, indexed by different values of wave vector along $z$. It is then possible to introduce the new second quantized operators, $\hr^\dagger_{\mathbf{q}}$ and $\hr_{\mathbf{q}}$, corresponding to the particular linear superposition of phonon modes that are coupled to electronic transitions. Based on reference~\cite{AP3_325(1954), PRB85_125302(2012)}, we can write the Hamiltonian as
\begin{align}
\hH_\text{phn-e} &= \sum_{\alpha,\mathbf{q}} \hbar \sqrt{\frac{\nu_\text{phn} e^2 \Delta n_{\alpha} }{4 \hbar \varepsilon_0 \varepsilon_\rho } \frac{\omega_\alpha}{\tilde{\omega}_{\alpha,\mathbf{q}}} \frac{I_{\alpha\alpha} \pc{q}}{\md{\mathbf{q}}}} \pc{\hb^\dagger_{\alpha, \mathbf{q}}+ \hb_{\alpha, -\mathbf{q}}} \pc{\hr_{-\mathbf{q}}^\dagger + \hr_{\mathbf{q}}},
\end{align}
with $\varepsilon_\rho^{-1} = \pc{\varepsilon_\infty^{-1} - \varepsilon_s^{-1}}^{-1}$.
Using the relations 
$$ \pc{\hb^\dagger_{\alpha,\mathbf{q}}  + \hb_{\alpha,-\mathbf{q}} } =\sqrt{ \frac{\omega_{\alpha}}{\tilde{\omega}_{\alpha,\mathbf{q}}}}\pc{ \hp^\dagger_{\alpha,\mathbf{q}}+ \hp_{\alpha,-\mathbf{q}} },$$
and
$$
\pc{ \hp^\dagger_{\alpha,\mathbf{q}}+ \hp_{\alpha,- \mathbf{q}} }= \sum_n X_{\alpha n} \pc{ \hP^\dagger_{n,\mathbf{q}} + \hP_{n,-\mathbf{q}}}
$$
where $X_{\alpha n} = \pc{x_{n \alpha} + y_{n \alpha}}^{-1}$, one can write the electron--phonon coupling in terms of the MSB plasmon operators as
\begin{align}
\hH_\text{phn-e}
 &= \sum_{\mathbf{q}} \sum_{\alpha,n} \hbar \sqrt{\frac{\nu_\text{phn} e^2 \Delta n_{\alpha} }{4 \hbar \varepsilon_0 \varepsilon_\rho } \frac{I_{\alpha\alpha} \pc{q}}{\md{\mathbf{q}}}  \frac{\omega_{\alpha}}{\tilde{\omega}_{\alpha,\mathbf{q}}} } X_{\alpha,n} \pc{ \hP^\dagger_{n,\mathbf{q}}+ \hP_{n,- \mathbf{q}} } \pc{\hr_{-\mathbf{q}}^\dagger + \hr_{\mathbf{q}}}.
\end{align}

\section{Exact Diagonalization of the Full~Hamiltonian \label{Sec:diag}}

Consider the Hamiltonian of one MSB plasmon coupled with a single phonon
\begin{align}
\hH &=\hH_\text{plm} + \hH_\text{phn} + \hH_\text{plm-phn}, \nonumber\\
&=\sum_\mathbf{q} \hbar \, W_\mathbf{q} \hP_\mathbf{q}^\dagger \hP_\mathbf{q} + \sum_\mathbf{q}  \hbar\, \nu_\text{phn} \hr^\dagger_\mathbf{q} \hr_\mathbf{q} + \sum_\mathbf{q}  \hbar\, \mathcal{G}_{\text{int},\mathbf{q}} \pc{\hP_\mathbf{q}^\dagger + \hP_\mathbf{-q}} \pc{\hr_{-\mathbf{q}}^\dagger + \hr_{\mathbf{q}}}
\end{align}
where $\mathcal{G}_{\text{int},\mathbf{q}}$ expresses the coupling between the phonons and the electronic excitation,
\begin{align}
\mathcal{G}_{\text{int},\mathbf{q}} = \sum_\alpha \sqrt{\frac{\nu_\text{phn} e^2 \Delta n_{\alpha} }{4 \hbar \varepsilon_0 \varepsilon_\rho } \frac{\omega_\alpha}{\tilde{\omega}_\alpha} \frac{I_{\alpha} \pc{q}}{ \md{\mathbf{q}}}} X_{\alpha \mathbf{q}}.
\label{eq:Gint}
\end{align}

Note that the anti-resonant terms of the phonon--MSB plasmons are included in the expression of the full Hamiltonian. Indeed these terms are important when the coupling strength is a significant fraction of the frequency of the individual oscillators, i.e.,~in the ultra-strong coupling regime~\cite{ciuti_PRB2005}.

We look for new coupled operators $\hq_1$ and $\hq_2$ such that
\begin{align}
\hq_{1,\mathbf{q}} &= a_{11} \hr_{\mathbf{q}} + b_{11} \hr_{-\mathbf{q}}^\dagger + a_{12} \hP_{-\mathbf{q}} + b_{12} \hP_{\mathbf{q}}^\dagger, \\
\hq_{2,\mathbf{q}} &= a_{21} \hr_{\mathbf{q}} + b_{21} \hr_{-\mathbf{q}}^\dagger + a_{22} \hP_{-\mathbf{q}} + b_{22} \hP_{\mathbf{q}}^\dagger.
\end{align}

The Hopfield coefficients introduced here satisfy the normalization condition
\begin{align*}
\md{a_{11}}^2 - \md{b_{11}}^2 + \md{a_{12}}^2-\md{b_{12}}^2 &= 1,\\
\md{a_{21}}^2 - \md{b_{21}}^2 + \md{a_{22}}^2-\md{b_{22}}^2 &= 1
\end{align*}
such that the Hamiltonian can be diagonalized as
\begin{align}
\hH = \hbar \Omega_{1,\mathbf{q}} \hq_{1,\mathbf{q}}^\dagger \hq_{1,\mathbf{q}} + \hbar \Omega_{2,\mathbf{q}} \hq_{2,\mathbf{q}}^\dagger \hq_{2,\mathbf{q}}.
\end{align}

Following a Hopfield--Bogoliubov diagonalization procedure, one obtains a system of linear equations~\cite{PR112_1555(1958)}. 
The eigenproblem for $n=1,2$ to find the eigenvalues $\Omega_{1}$ and $\Omega_{2}$ is
\begin{align}
\Omega_{n} 
\left( \begin{array}{c}
a_{n1}\\ b_{n1} \\ a_{n2} \\ b_{n2} \end{array} \right) =
\left( \begin{array}{cccc}
\nu_\text{phn} & 0 & \mathcal{G}_{\text{int},\mathbf{q}} & - \mathcal{G}_{\text{int},\mathbf{q}} \\ 
0 & -\nu_\text{phn} & \mathcal{G}_{\text{int},\mathbf{q}} & - \mathcal{G}_{\text{int},\mathbf{q}} \\ 
\mathcal{G}_{\text{int},\mathbf{q}} & -\mathcal{G}_{\text{int},\mathbf{q}} & W_\mathbf{q} & 0 \\ 
\mathcal{G}_{\text{int},\mathbf{q}} & -\mathcal{G}_{\text{int},\mathbf{q}}& 0 & - W_\mathbf{q} \\ 
 \end{array} \right) 
\left( \begin{array}{c}
a_{n1}\\ b_{n1} \\ a_{n2} \\ b_{n2} \end{array} \right) .
\end{align}

The solutions of this system of equations are two MSB polaron modes. The~characteristic equation of the eigenproblem ($\det \pr{I\lambda - M}=0$) can be written as 
\begin{align}
\left(\nu_\text{phn} ^2-\lambda ^2\right) \left(W_\mathbf{q}^2-\lambda ^2\right)-4 \nu_\text{phn}  W_\mathbf{q}  \mathcal{G}_{\text{int},\mathbf{q}}^2  =0.
\end{align}

The solutions of the above equation are given by
\begin{align}
\Omega_\text{UP}^2 &= \frac{1}{2} \pc{\nu_\text{phn} ^2+W_\mathbf{q} ^2 + \sqrt{ \pc{\nu_\text{phn}^2-W_\mathbf{q}^2}^2 + 16 \mathcal{G}_{\text{int},\mathbf{q}}^2 \nu_\text{phn}  W_\mathbf{q}}},\\
\Omega_\text{LP}^2 &= \frac{1}{2} \pc{\nu_\text{phn} ^2+W_\mathbf{q} ^2 - \sqrt{ \pc{\nu_\text{phn}^2-W_\mathbf{q}^2}^2 + 16 \mathcal{G}_{\text{int},\mathbf{q}}^2 \nu_\text{phn}  W_\mathbf{q}}}.
\end{align}

The Hopfield coefficients can be expressed in closed form. We can define a phononic part $h_\text{phn} = \md{a_{n1}}^2 - \md{b_{n1}}^2$ and a plasmonic part $h_\text{plm} = \md{a_{n2}}^2- \md{b_{n2}}^2$ linked by the relation $h_\text{phn} + h_\text{plm} = 1$.
For the phononic part, we obtain the expressions
\begin{align}
h_\text{phn}^\text{UP} &= \frac{\Omega_\text{UP}^2 - W_\mathbf{q}^2}{\Omega_\text{UP}^2-\Omega_\text{LP}^2}, \\
h_\text{phn}^\text{LP} &= \frac{W_\mathbf{q}^2 -\Omega_\text{LP}^2 }{\Omega_\text{UP}^2-\Omega_\text{LP}^2} .
\label{eq:hopfield}
\end{align}

Note that we have necessarily $h_\text{phn}^\text{UP} + h_\text{phn}^\text{LP}= 1$ (and therefore $h_\text{plm}^\text{UP} + h_\text{plm}^\text{LP}= 1$).

As a linear superposition of a MSB plasmon and a phonon, the~lifetime of the MSB polarons is directly determined by the decay rate of the MSB plasmon, $\gamma_\text{ISB}$, and~the phonon decay rate, $\gamma_\text{phn}$, as
\begin{align}
\gamma_\text{LP} &= \md{h_\text{phn}^\text{LP}}^2 \gamma_\text{phn} +\md{h_\text{ISB}^\text{LP}}^2 \gamma_\text{ISB},\\
\gamma_\text{UP} &= \md{h_\text{phn}^\text{UP}}^2 \gamma_\text{phn} +\md{h_\text{ISB}^\text{UP}}^2 \gamma_\text{ISB}.
\end{align}

\section{Discussion and~Results \label{Sec:results}}

\begin{figure}[h]
\centering
\includegraphics[width=0.55\linewidth]{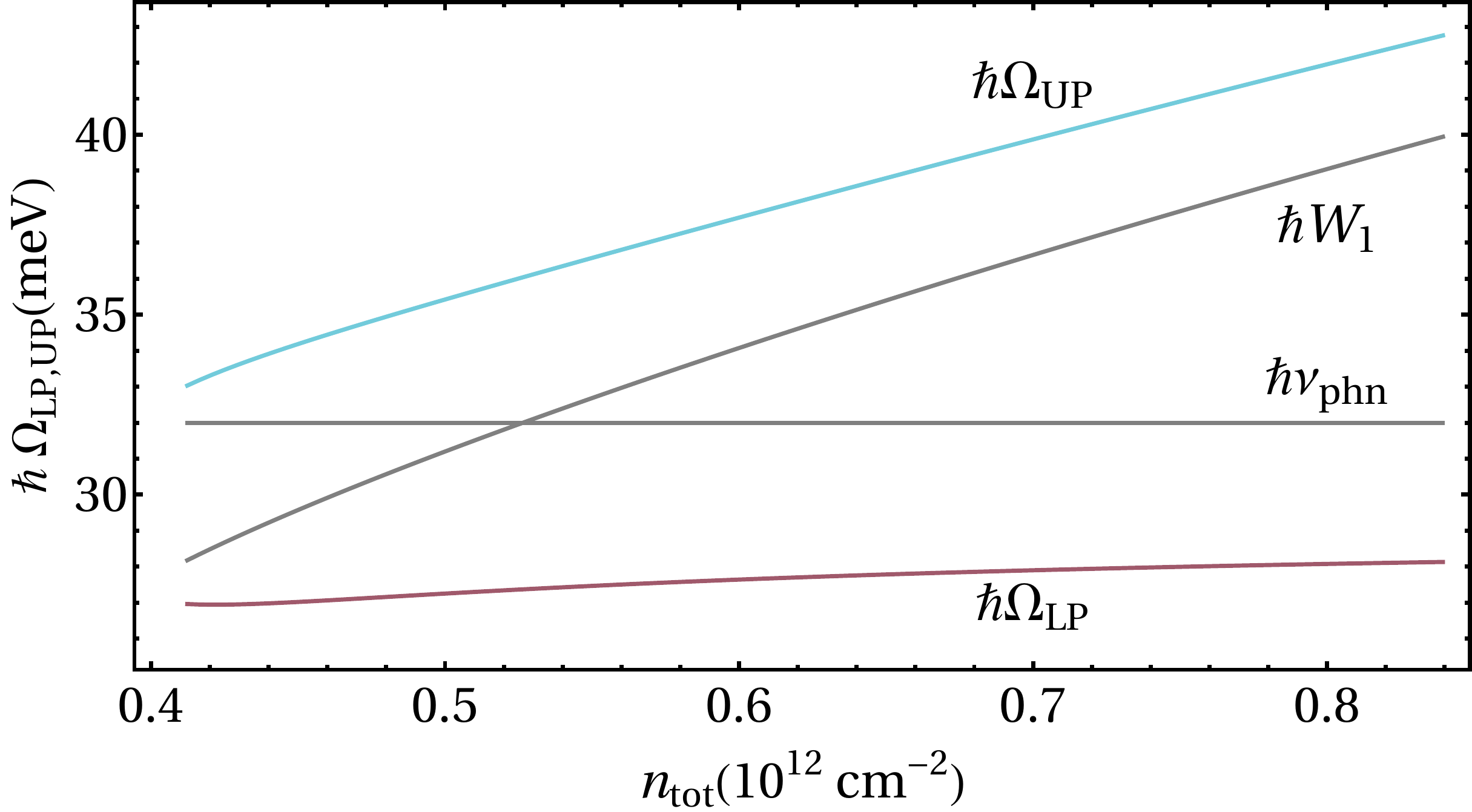}\\
\caption{(Color online) Phonon and MSB  free energies ($\hbar W_{1}$ and $\hbar \nu_\text{phn}$) and coupled eigenmodes ($ \hbar \Omega_\text{LP,UP}$, in~purple and blue) as a function of the total doping density $n_\text{tot}$ in the quantum~well. \label{Fig:MSBphn1}}
\vspace*{0.5cm}
\includegraphics[width=0.55\linewidth]{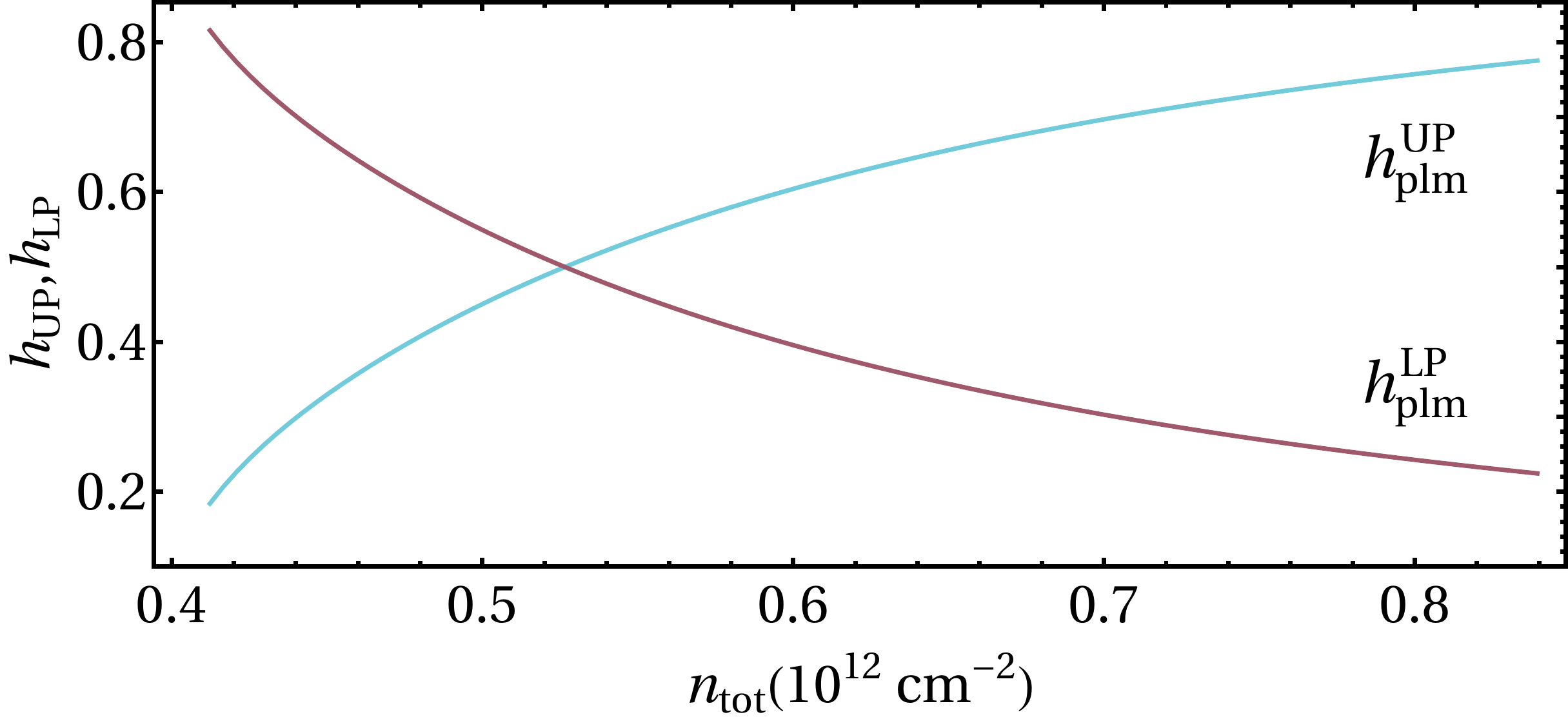}\\
\caption{(Color online) Hopfield coefficients Equations~\eqref{eq:hopfield} as a function of the total doping density $n_\text{tot}$ in the quantum~well.\label{Fig:Hopfield}}
\end{figure}

In Figure~\ref{Fig:MSBphn1}, we took the same system as in Figure~\ref{Fig:MSBexample}, and~plotted the MSB polaron frequencies $\Omega_\text{LP, UP}$ as a function of the total electron density in the quantum well. The~frequencies of the LO phonon ($\hbar \nu_{\text{phn}}$ = 32~meV) and MSB plasmon modes are plotted on the same graph for better understanding. The~resonance condition $W_1 = \nu_\text{phn}$ is verified at $n_\text{tot} \approx 5.26 \times 10^{11}$~cm$^{-2}$ where $\hbar \mathcal{G}_\text{int} \approx 4.3$~meV. 

One can see that the strength of the interaction between LO phonons and MSB plasmon leads to the shifting of the phonon and the MSB energies and mixing or hybridization between the two modes -- the coupling produces a {repulsion} between the two free modes in the region $W_1 = \nu_\text{phn}$.
While one might not recognize easily the typical avoided crossing curves for this type of problems, both curves converge, as~expected, in~the limits of small $n_\text{tot}$ to the uncoupled modes frequencies, that is, $\Omega_\text{UP}$ converges to $\nu_\text{phn}$ and $\Omega_\text{LP}$ to $\omega_{23}$ (see the limits of $W_1$ in Figure~\ref{Fig:MSBexample}).
It is particularly interesting to have the possibility to control the MSB plasmon--phonon hybridization by varying the doping density of the quantum well.
As such, an~interesting aspect to study is the mixing between the phononic and electronic degrees of freedom. The~Hopfield coefficients defined in Equation~\eqref{eq:hopfield} quantify this mixing and are plotted in Figure~\ref{Fig:Hopfield}. The~maximum mixing occurs as expected when $W_1 = \nu_\text{phn}$. 

In Table~\ref{tablecoupling}, we show the results for normalized coupling at $W_\text{MSB} = \nu_\text{phn}$ for different materials. The~different values for $n_\text{tot}$ are chosen considering the fact that only the first two subbands of the quantum well are occupied and maximizing the coupling ratio. 
It is clear that the coupling strength depends strongly on the levels of doping that is possible to achieve, but~also on the different materials parameters. From~the chosen materials, we can observe that the combination of higher effective electron mass and phonon energy together with smaller relative permittivity has a tendency to lead to stronger couplings (see Equation~\eqref{eq:Gint}).
One can infer that, same as for intersubband polarons~\cite{PRB85_125302(2012)}, the~origin of such large coupling strengths can be found in the superradiant nature of MSB excitations and in the natural confinement of the phonons inside the quantum well. From~this table it is possible to see that the MSB plasmon--LO-phonon coupling can be up to $40\%$ of the phonon frequency, indicating that the system is in the ultra-strong coupling regime. This relative coupling energy could be even more important, by~increasing the electronic density and the number of occupied subbands. The~existence of a strong interaction between plasmons and LO-phonons could lead to the investigation of processes of stimulated scattering of plasmons mediated by phonons, in~analogy to what has been done with intersubband polaritons~\cite{PRL102_136403(2009)}.
\begin{table}
\centering
\caption{Normalized coupling at resonance $W_\text{MSB} = \nu_\text{phn}$ as a function of the longitudinal optical (LO) phonon frequency for different materials~\cite{Parameters} \label{tablecoupling}.}
 \begin{tabular}{|l |c c c|}
 \hline 
& $\boldsymbol{n_\textbf{tot}}$~$\boldsymbol{\pc{\textbf{cm}^{-2}}}$ & $\boldsymbol{\hbar \nu_\textbf{phn}}$\textbf{~(meV)} & $\boldsymbol{\mathcal{G}_\textbf{int} / \nu_\textbf{phn}}$ \\
\hline
InAs & $3.7 \times 10^{11}$ & 30.0 & 0.161  \\
GaInAs & $6.5 \times 10^{11}$ & 32.0 & 0.175 \\
GaAs & $9.6 \times 10^{11}$ & 35.0 & 0.171   \\
InP & $1.4 \times 10^{12}$ & 43.0 & 0.218  \\
CdTe & $6.5 \times 10^{11}$ & 21.0 & 0.282  \\
ZnSe & $1.3 \times 10^{12}$ & 31.0 & 0.304  \\
GaN & $3.8 \times 10^{12}$ & 87.3 & 0.332  \\
ZnO & $3.8 \times 10^{12}$ & 72.0 & 0.395  \\
\hline
\end{tabular}
\label{Table_parameters}
\end{table}

\section{Conclusions \label{Sec:conclusions}}
The diagonalization of the complete MSB plasmon--phonon interaction Hamiltonian provided us with the mixed excitations of the coupled system denominated MSB polarons. The~coupling energy has been calculated for different material systems. We demonstrated that a ratio of almost $40\%$ between the coupling and the phonon frequency can be achieved in ZnO material system. Even higher values can be achieved with MSB plasmons issued from quantum wells with several occupied~subbands. 

In this work we only took into account longitudinal optical phonons, which are not coupled with light. A~natural perspective of this work is the study of the coupling of both plasmons and phonon modes to electromagnetic radiation. For~this, one has to redefine a matter polarization including both transverse and optical phonon modes, which has to be added to the contribution of the intersubband polarization. For~this purpose recent work by Francki\'e~et~al.~\cite{PRB97_075402(2018)} could be exploited. They have studied a quantum cascade phonon polariton laser based on an inverted intersubband transition weakly coupled to a quasi-particle issued from the strong coupling between a transverse optical (TO) phonon and a cavity mode. For~their study, they have introduced a polarization operator associated with the TO-phonons in the Power--Zienau--Woolley gauge. The~diagonalization of the Hamiltonian including phonons, intersubband excitations and electromagnetic field on the same ground will allow one to accurately describe how the material Reststrahlen band can be modified by the presence of collective electronic excitations. Another interesting perspective, which goes beyond the scope of this work, concerns the investigation of the impact of a modified interaction with phonons induced by collective effects on the thermal conductivity of the doped semiconductor. Indeed it has been shown experimentally and theoretically that electron--phonon interaction can modify both the electronic and lattice contribution to the thermal conductivity~\cite{APL115_023903(2019), physrev182}.

\subsection*{Acknowledgments}
We acknowledge financial support from Labex SEAM.


\end{document}